\documentclass{article}

\usepackage{graphicx}
\usepackage{xspace}
\usepackage{amsmath}

 \usepackage[dblblindworkshop, final]{neurips_2026}
  \workshoptitle{SaTQuML: Secure and Trustworthy Quantum Machine Learning}

\usepackage[utf8]{inputenc} 
\usepackage[T1]{fontenc}    
\usepackage{hyperref}       
\usepackage{url}            
\usepackage{booktabs}       
\usepackage{amsfonts}       
\usepackage{nicefrac}       
\usepackage{microtype}      
\usepackage{xcolor}         
\usepackage{algorithm}
\usepackage{algpseudocode}
\algrenewcommand\algorithmicrequire{\textbf{Input:}}
\algrenewcommand\algorithmicensure{\textbf{Output:}}
\usepackage{subcaption}
\usepackage{braket}

\title{On the Impact of Coordinate Descent for Multi-Angle QAOA in the Independent Set Graph Problem}

\workshoptitle{SaTQuML: Secure and Trustworthy Quantum Machine Learning}
\author{%
  Daeyeun Kim \\
  Korea University \\
  Seoul, Republic of Korea \\
  \texttt{dk725@korea.ac.kr}
  \And
  Seungcheol Oh \\
   Korea University \\
  Seoul, Republic of Korea \\
  \texttt{seungoh@korea.ac.kr}
  \And
  Joongheon Kim\thanks{Joongheon Kim is the corresponding author. This research was supported by IITP (Institute for Information \& Communications Technology Planning \& Evaluation) funded by MSIT (Ministry of Science and ICT), Korea, (RS-2024-00439803, SW Star Lab) for Quantum AI Empowered Second-Life Platform Technology.} \\ 
  Korea University \\
  Seoul, Republic of Korea \\
  \texttt{joongheon@korea.ac.kr}
}

\begin{document}

\maketitle

\begin{abstract}
Constrained combinatorial optimisation provides a mathematical framework for modelling a wide range of practical decision making problems, including energy network operation, financial portfolio optimisation, logistics, routing, and resource allocation. This paper studies parameter optimisation for the multi angle quantum approximate optimisation algorithm (maQAOA) applied to the maximum independent set problem, a canonical constrained graph based optimisation task. We propose a coordinate wise training framework for maQAOA that updates one variational parameter at a time by solving a sequence of one dimensional optimisation subproblems. For each selected coordinate, the method reconstructs the objective dependence on that parameter and moves directly to the parameter value that minimises the training objective, or equivalently maximises the corresponding expected solution quality. While demonstrated specifically on maQAOA, the underlying theoretical principle, that the expectation value with respect to a given parameter can be analytically expressed as a finite Fourier series, is broadly applicable to general tasks solved via parameterised quantum circuits. Empirical evaluations on connected Erdős Rényi graph benchmarks demonstrate that this analytic approach significantly reduces the computational training cost required to reach target approximation ratios and yields superior convergence reliability compared to standard gradient based, stochastic, and derivative free baselines.
\end{abstract}

\section{Introduction}
\textbf{Background and Motivation.}
The earliest practically useful quantum processors are expected to operate in the noisy intermediate scale quantum (NISQ) regime~\citep{preskill2018quantum}. In this regime, limited qubit counts, finite coherence times, and imperfect gate fidelities restrict the size and depth of executable circuits. Overcoming these hardware limitations to solve hard combinatorial optimisation problems remains a central goal in quantum computing and algorithms. These hardware constraints have motivated hybrid methods combining quantum and classical computation, in which the quantum processor is used to prepare and measure parameterised quantum states, while a classical routine performs the outer loop optimisation. Variational quantum algorithms (VQAs) are a central example of this paradigm: a parameterised quantum circuit (PQC) is evaluated on a quantum processing unit, and its parameters are updated by a classical optimiser using objective values derived from measurements~\citep{cerezo2021variational,mcclean2016theory}.

Among VQAs, the quantum approximate optimisation algorithm (QAOA) is a prominent framework that leverages PQCs to solve combinatorial optimisation problems~\citep{farhi2014qaoa,blekos2024review}. QAOA is specifically designed for discrete optimisation, a paradigm that encompasses many practical decision making tasks through quadratic unconstrained binary optimisation (QUBO) or Ising formulations~\citep{lucas2014ising,glover2022quantum}. Such formulations are deeply intertwined with Boolean satisfiability and constraint logic, as they mathematically map rigid logical requirements into trainable energetic penalties. These formulations encompass canonical problem classes such as covering~\citep{lucas2014ising}, packing~\citep{martello1987algorithms}, assignment~\citep{kuhn1955hungarian}, scheduling~\citep{lawler1993sequencing}, and routing~\citep{dantzig1959truck}. Important practical applications explored in current QAOA literature include logistics~\citep{xie2024feasibility,ruan2020quantum}, finance~\citep{venturelli2019reverse}, and communications~\citep{apwcs21}. 

To achieve high quality solutions in these complex application domains, the underlying quantum circuit must possess sufficient representational capacity. In standard QAOA, this capacity is traditionally scaled by deepening the circuit architecture; specifically, increasing the number of alternating cost and mixer layers generally improves the expressive power of the circuit and can lead to better approximation quality. On NISQ hardware, however, this route is severely limited by noise accumulation due to the aforementioned hardware constraints, making high depth QAOA difficult to execute without the desired information being destroyed by noise~\citep{blekos2024review}. One prominent solution to this challenge is multi-angle QAOA (ma-QAOA)~\citep{herrman2022multi}. By assigning an independent variational parameter to each individual quantum gate, the required circuit depth can be substantially reduced for a given level of approximation quality. This architectural adjustment effectively shifts the computational burden from the fragile quantum hardware to the classical parameter optimisation routine. However, the resulting increase in parameter dimensionality drastically enlarges the optimisation landscape, making the classical search for an optimal set of parameters more difficult and increasingly susceptible to classical inefficiencies.

\noindent \textbf{Algorithm Concept.}
To efficiently optimise the enlarged ma-QAOA parameter space, this paper proposes a cyclic coordinate descent strategy akin to sequential minimal optimisation and Rotosolve~\citep{nakanishi2020sequential,ostaszewski2021structure}. This approach explicitly leverages the analytical properties of the quantum loss landscape to bypass noisy gradient estimations in high dimensions. Since each ma-QAOA angle is associated with a single rotation gate, the one dimensional objective is purely sinusoidal~\citep{wierichs2022general}, enabling analytical coordinate updates from just a few expectation value evaluations.

\noindent \textbf{Contributions.}
This paper investigates cyclic coordinate descent as a training method for ma-QAOA, offering a practical pathway to scale techniques utilising PQCs for complex combinatorial tasks. Although we apply this cyclic coordinate descent specifically to ma-QAOA, the theoretical foundation of the method, namely that the expectation value with respect to a chosen parameter can be analytically expressed as a finite Fourier series, is broadly applicable for optimising parameters across general tasks and problems solved via PQCs. We first explain why the cyclic coordinate descent is particularly compatible with ma-QAOA, highlighting how it mitigates the curse of dimensionality introduced by independent variational angles. We then compare its empirical performance against standard optimisation methods commonly used for ma-QAOA and broader PQC training. Finally, using the maximum independent set problem, a canonical NP hard task fundamentally linked to constrained satisfiability formulations, as a constrained combinatorial optimisation benchmark, we demonstrate that the proposed method achieves significantly more efficient parameter optimisation relative to the baselines considered.

\section{QAOA on Constrained Optimisation Problems}

\subsection{Overview of QAOA}
QAOA can be regarded as a variational, gate-based analogue of adiabatic quantum computation~\citep{farhi2014qaoa,blekos2024review,albash2018adiabatic}. In standard adiabatic computation, a quantum system is prepared in an easy-to-construct eigenstate of an initial Hamiltonian and then evolved slowly towards a more complex problem Hamiltonian. Under appropriate spectral conditions, specifically if the evolution time is sufficiently long relative to the minimum energy gap between the ground and first excited states, the adiabatic theorem ensures that the state remains close to the corresponding instantaneous eigenstate during the continuous time evolution~\citep{albash2018adiabatic}. The initial Hamiltonian is intentionally chosen so that its relevant extremal eigenstate is easy to prepare, while the problem Hamiltonian is engineered to possess an extremal eigenstate encoding the desired classical solution.

\begin{figure}[t!]
    \centering
    \hspace{0.04\columnwidth}
    \includegraphics[width=0.65\columnwidth]{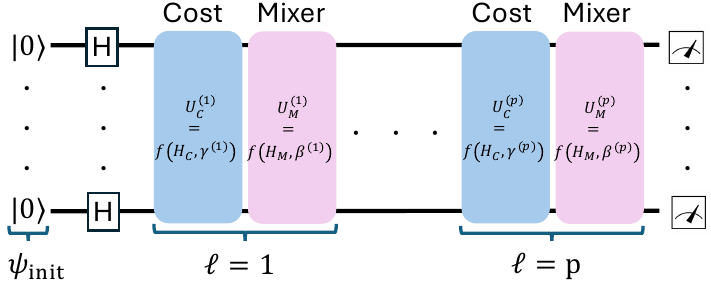}
    \caption{Circuit illustration of QAOA algorithm.}
    \label{fig:QAOAcircuit}
    \vspace{-3mm}
\end{figure}

QAOA replaces this continuous analog evolution with a discretised, finite sequence of alternating unitaries generated by a cost Hamiltonian $H_C$ and a mixer Hamiltonian $H_M$, leading to the layered circuit structure illustrated in Fig.~\ref{fig:QAOAcircuit}. Typically, the cost Hamiltonian $H_C$ is constructed to be diagonal in the computational basis, mapping classical bitstrings directly to their objective function values, while the mixer Hamiltonian $H_M$ generally consists of non-commuting transverse field operators that drive transitions between these basis states. To initialise the standard unconstrained QAOA circuit into the maximal eigenstate of this mixer, a Hadamard gate is applied to each qubit of the state $\ket{\psi_{\text{init}}} = \ket{0}^{\otimes n}$. This creates the uniform superposition state $\ket{+}^{\otimes n}$ across all possible computational basis states, with $n$ representing the total qubit count. 

Subsequently, a depth-$p$ parameterised ansatz is constructed by applying $p$ layers of these alternating operators to yield the final quantum state before measurement:
\begin{equation}
    \ket{\psi_p(\gamma,\beta)} =
    \left( \prod_{\ell=1}^{p}
    e^{-i\beta_\ell H_M}
    e^{-i\gamma_\ell H_C}\right)
    \ket{+}^{\otimes n},
    \label{eq:qaoa-state}
\end{equation}
with continuous variational parameters $\gamma=(\gamma_1,\dots,\gamma_p)$ controlling the phase separation generated by the cost Hamiltonian, and $\beta=(\beta_1,\dots,\beta_p)$ controlling the interference generated by the mixer Hamiltonian. 

After the circuit is executed and measured, the expectation value of the cost Hamiltonian,
\begin{equation}
    F_p(\gamma,\beta) =
    \bra{\psi_p(\gamma,\beta)}{H_C}\ket{\psi_p(\gamma,\beta)}
    \label{eq:qaoa-objective}
\end{equation}
is estimated through repeated state preparations and projective measurements. This expectation value acts as a parameterised loss or reward landscape and is used by a classical optimiser in an outer-loop procedure to iteratively update $(\gamma,\beta)$. The objective is maximised or minimised depending on the specific formulation of the optimisation problem. 

For a maximisation problem, given optimised parameters $(\gamma^\star,\beta^\star)$ returned by the classical routine, the algorithm's performance is commonly evaluated using the approximation ratio,
\begin{equation}
    \alpha_p =
    \frac{F_p(\gamma^\star,\beta^\star)}{C_{\max}},
    \label{eq:qaoa-approximation-ratio}
\end{equation}
where $C_{\max}$ is the maximum objective value attainable. While standard adiabatic computing requires infinitely slow evolution to guarantee optimality, it is theoretically established that as $p \to \infty$, QAOA recovers the adiabatic guarantee and yields the optimal solution~\citep{farhi2014qaoa}. However, for near-term implementation at any finite depth $p$, QAOA effectively reduces the computationally intractable search over an exponentially large space of $2^n$ discrete bit strings to the continuous, heuristic optimisation of only $2p$ parameters. Because this parameter dimensionality depends strictly on the chosen circuit depth and remains entirely decoupled from the size of the underlying problem instance, QAOA serves as a highly scalable variational framework.

\subsection{Constraint-Handling Strategies}
Standard QAOA is naturally suited to unconstrained binary optimisation, since the mixer explores the full bitstring space. For constrained problems, only a subset \(\mathcal{F}\subseteq\{0,1\}^n\) is feasible, so additional mechanisms are needed to prevent or discourage infeasible samples. A common approach is to absorb the constraints into the objective using penalty terms, thereby converting the problem into an unconstrained QUBO or Ising formulation~\citep{blekos2024review,lucas2014ising,glover2022quantum,hadfield2017quantum}. For a maximisation problem, this can be written as 
\begin{equation}
C_{\mathrm{pen}}(x)=C(x)-\lambda V(x),     
\end{equation}
where \(V(x)=0\) for feasible strings and \(V(x)>0\) otherwise. This method is simple and compatible with standard QAOA, but its performance depends strongly on the penalty weight \(\lambda\): weak penalties may favour infeasible high-value states, while overly large penalties can dominate the original objective and complicate optimisation. Alternatively, penalty-free approaches enforce feasibility by construction, such as in constrained QAOA or the quantum alternating operator ansatz~\citep{hadfield2017quantum,hadfield2019quantum}. In this setting, the initial state is chosen from \(\mathcal{F}\), and the mixer is designed to map feasible states only to feasible states. Consequently, unlike the standard QAOA initialisation, hard-constrained variants do not generally apply Hadamard gates to all qubits at the start of the circuit because applying \(H^{\otimes n}\) would populate infeasible bitstrings. Examples include \(XY\)-type mixers for cardinality constraints and problem-specific controlled mixers for independent-set problems~\citep{wang2020xy,saleem2020max,fuchs2022constraint,bartschi2020grover,xie2024feasibility}. Such mixers avoid penalty tuning and ensure feasible samples, but they require problem-dependent circuit design and may increase depth~\citep{bucher2025penalty}.

\section{Coordinate Descent for Multi-Angle QAOA}
\subsection{Overview of Multi-Angle QAOA}
In standard QAOA, each layer is controlled by a single cost angle and a single mixer angle, shared across all corresponding gates in that layer. By contrast, ma-QAOA assigns independent variational angles to the individual gates in cost and mixer blocks~\citep{herrman2022multi}. For instance, in the MaxCut problem on a graph with $n=|V|$ vertices and $m=|E|$ edges, the cost and mixer blocks in each layer require $m$ and $n$ gates, respectively. Correspondingly, the total number of parameters increases from $2p$ in standard QAOA to $(m+n)p$ in ma-QAOA. Standard QAOA thus emerges as a specific instance of ma-QAOA where the parameters remain uniform across each layer. Because it strictly generalises the standard approach, ma-QAOA guarantees that its optimal objective value at any given depth $p$ is at least as good as that of standard QAOA~\citep{herrman2022multi}. Empirical studies have shown that ma-QAOA can match or exceed standard-QAOA approximation ratios while using substantially fewer layers on a range of benchmark instances~\citep{herrman2022multi,shi2022multiangle,gaidai2024performance}. While ma-QAOA's flexibility is ideal for depth-limited NISQ devices, it severely burdens classical optimisation due to the increased number of parameters. To achieve real-world performance gains on near-term hardware, it is essential to develop efficient strategies to handle this expanded parameter space.

\subsection{Overview of Coordinate Descent Method}
For a broad class of PQCs, suppose all parameters except one free parameter \(\theta_{\text{free}}\) are held fixed, and that \(\theta_{\text{free}}\) enters the circuit through Pauli rotations of the form \(U(\theta_{\text{free}})=e^{-i\theta_{\text{free}}P/2}\), where \(P^2=I\). Then, the expectation value of any observable, viewed as a function of \(\theta_{\text{free}}\), admits a finite Fourier series representation whose degree scales linearly with the number of gates controlled by \(\theta_{\text{free}}\)~\citep{nakanishi2020sequential,ostaszewski2021structure,wierichs2022general}. This structure makes coordinate descent training particularly natural for ma-QAOA, since each variational parameter is associated with a single rotation gate. Consequently, the corresponding one-dimensional objective reduces to a single-frequency sinusoid, $f(\theta_{\text{free}}) = A\sin(\theta_{\text{free}}+\phi)+C$, for constants \(A\), \(\phi\), and \(C\). This sinusoidal form can be reconstructed from only three expectation-value evaluations, allowing the exact coordinate-wise extremum to be determined analytically. The full training procedure cyclically sweeps through all \(D\) ma-QAOA parameters, updating one coordinate at a time until either a maximum number of sweeps is reached or the improvement falls below a prescribed tolerance. A direct implementation requires \(3D\) expectation-value evaluations per sweep. By reusing the current objective value after each coordinate update, this cost can be reduced to \(2D+1\) evaluations for the first sweep and \(2D\) evaluations for each subsequent sweep. As a gradient-free approach, the method requires no hyperparameters to govern search dynamics; the only required parameters define the stopping criteria. The only hyperparameters to be specified are those to define stopping criteria. In the noiseless setting, each update moves the selected coordinate directly to the analytic extremum of its one-dimensional objective, so the objective is monotonically non-decreasing for maximisation tasks. However, because the method updates only one coordinate at a time, it does not explicitly capture correlations among parameters and may converge to a suboptimal coordinate-wise optimum. Algorithm~\ref{alg:cd-maqaoa} outlines the coordinate descent training procedure for ma-QAOA.

\begin{algorithm}[t!]
\caption{Cyclic coordinate descent for ma-QAOA}
\label{alg:cd-maqaoa}
\small
\begin{algorithmic}[1]
\Require ma-QAOA objective \(F_p(\Theta)\), initial parameter vector
\(\Theta^{(0)}=(\theta_1^{(0)},\ldots,\theta_D^{(0)})\), maximum number of sweeps \(T_{\max}\), tolerance \(\varepsilon\), optimisation sign \(s\in\{+1,-1\}\) 
\Ensure Optimised parameter vector \(\Theta^\star\)

\State \(\Theta \gets \Theta^{(0)}\)
\State \(F^{\mathrm{curr}} \gets F_p(\Theta)\) \Comment{expectation-value evaluation}

\For{\(t=1,\ldots,T_{\max}\)}
    \State \(F^{\mathrm{start}} \gets F^{\mathrm{curr}}\)

    \For{\(d=1,\ldots,D\)}
        \State \(\varphi \gets \theta_d\)
        \State \(f_0 \gets F^{\mathrm{curr}}\)
        \State \(f_+ \gets F_p(\theta_1,\ldots,\theta_d+\pi/2,\ldots,\theta_D)\) \Comment{expectation-value evaluation}
        \State \(f_- \gets F_p(\theta_1,\ldots,\theta_d-\pi/2,\ldots,\theta_D)\) \Comment{expectation-value evaluation}

        \State \(g \gets 2f_0-f_+-f_-\)
        \State \(h \gets f_+-f_-\)

        \State \(\theta_d^\star \gets
        \mathcal{W}\!\left(
        \varphi+s\frac{\pi}{2}
        -\operatorname{atan2}(g,h)
        \right)\) 
        \Comment{\(\mathcal{W}(\cdot)\) wraps the angle modulo \(2\pi\)}

        \State \(\theta_d \gets \theta_d^\star\)
        \State \(F^{\mathrm{curr}} \gets
        \dfrac{f_+ + f_-}{2}
        + s\,\dfrac{\sqrt{g^2+h^2}}{2}\)
    \EndFor

    \If{\(\left|F^{\mathrm{curr}}-F^{\mathrm{start}}\right| < \varepsilon\)}
        \State \textbf{break}
    \EndIf
\EndFor

\State \(\Theta^\star \gets \Theta\)
\State \Return \(\Theta^\star\)
\end{algorithmic}
\end{algorithm}

\section{Performance Evaluation}
\subsection{Evaluation Setup} 
We evaluate the proposed training method on the maximum independent set problem. For a graph \(G=(V,E)\), each vertex is represented by a single qubit, and a measured bit value \(x_i=1\) indicates that vertex \(i\) is selected; feasible solutions therefore satisfy \(x_i x_j=0\) for every edge \((i,j)\in E\). We generated the benchmark connected Erdős--Rényi graphs from the families in Table~\ref{tab:graphCases} using 15 random seeds each. Constraints are handled using a feasibility-preserving MIS mixer: the circuit starts from \(|0\rangle^{\otimes n}\), without initial Hadamard gates, and applies a controlled \(R_X\) rotation to a vertex only when all of its neighbouring vertices are unselected. We compare the proposed method with representative gradient-based, stochastic, and derivative-free baselines, including vanilla gradient descent (GD), adaptive moment estimation (Adam)~\citep{kingma2014adam}, simultaneous perturbation stochastic approximation (SPSA)~\citep{spall2002multivariate}, constrained optimisation by linear approximations (COBYLA)~\citep{powell1994direct}, the Nelder--Mead simplex method (NM)~\citep{nelder1965simplex}, Powell's derivative-free conjugate-direction method (Powell)~\citep{powell1964efficient}, and the limited-memory Broyden--Fletcher--Goldfarb--Shanno algorithm with bound constraints (L-BFGS-B)~\citep{byrd1995limited}. Since optimisers differ in the number of objective calls required per update, we report training cost in terms of the number of expectation-value evaluations rather than the number of optimiser iterations. Each expectation-value evaluation corresponds to estimating the objective value at a fixed parameter setting; on quantum hardware, this requires repeated circuit preparations and measurements, with the total shot cost scaling with the chosen shot budget. This metric therefore provides a hardware-relevant proxy for the computational cost incurred on the quantum device.

\subsection{Evaluation Results}
\begin{figure*}[h!]
    \centering
    \begin{subfigure}[t]{0.98\textwidth}
        \centering
        \includegraphics[width=\linewidth]{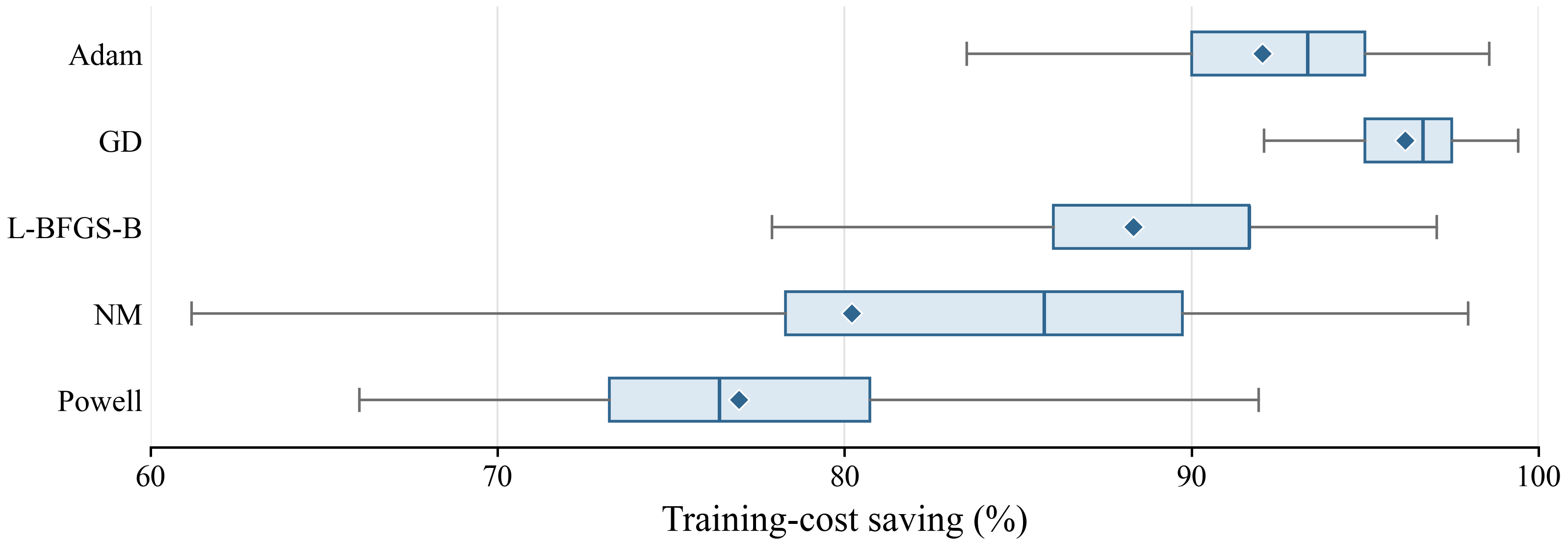}
        \caption{}
        \label{fig:fiveOptimiser_box}
    \end{subfigure}
    \hfill
    \begin{subfigure}[t]{0.98\textwidth}
        \centering
        \includegraphics[width=\linewidth]{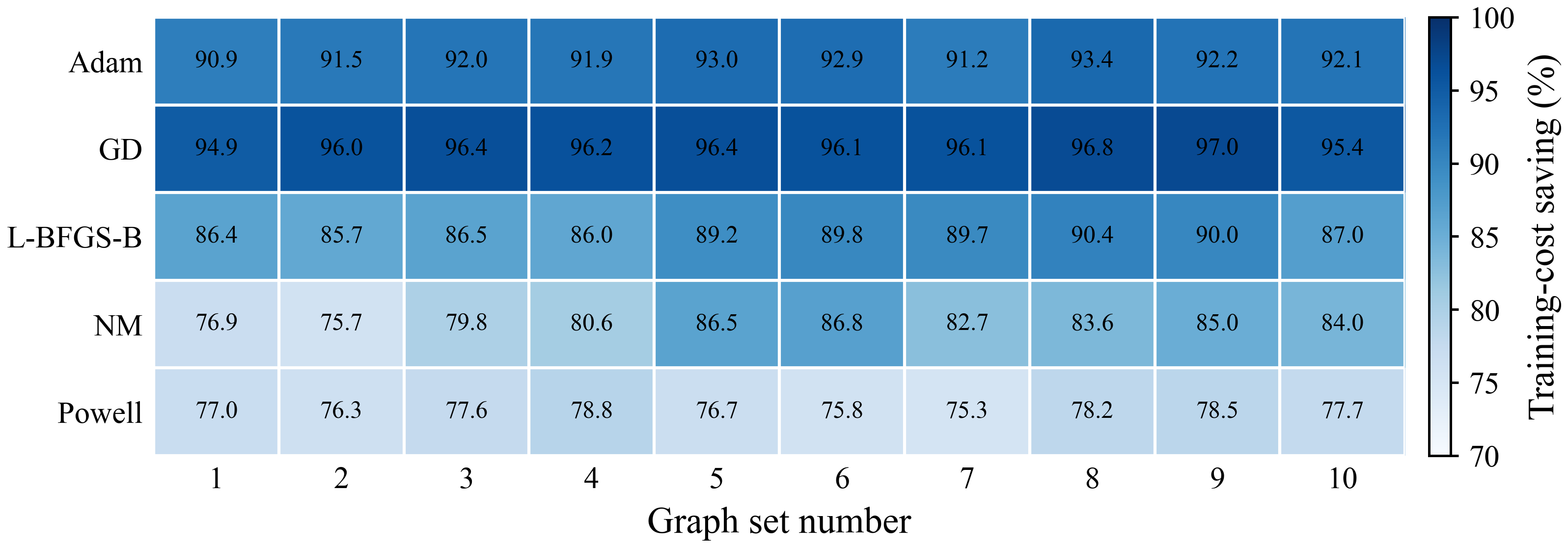}
        \caption{}
        \label{fig:fiveOptimiser_heat}
    \end{subfigure}
    \caption{Percentage savings in training cost achieved by the proposed method relative to baseline optimisation algorithms (Adam, GD, L-BFGS-B, NM, Powell) for crossing a target approximation ratio of 0.94. The results are presented as (a) box-and-whisker plots representing the distribution of savings across individual evaluation cases, where diamond markers indicate the overall mean savings (GD: 96.2\%, Adam: 92.0\%, L-BFGS-B: 88.3\%, NM: 80.2\%, Powell: 77.0\%), and (b) a heatmap detailing the average savings per graph set.}
    \label{fig:fiveOptimiser}
\end{figure*}

\begin{figure*}[h!]
    \centering
    \begin{subfigure}[t]{0.439\textwidth}
        \centering
        \includegraphics[width=\linewidth]{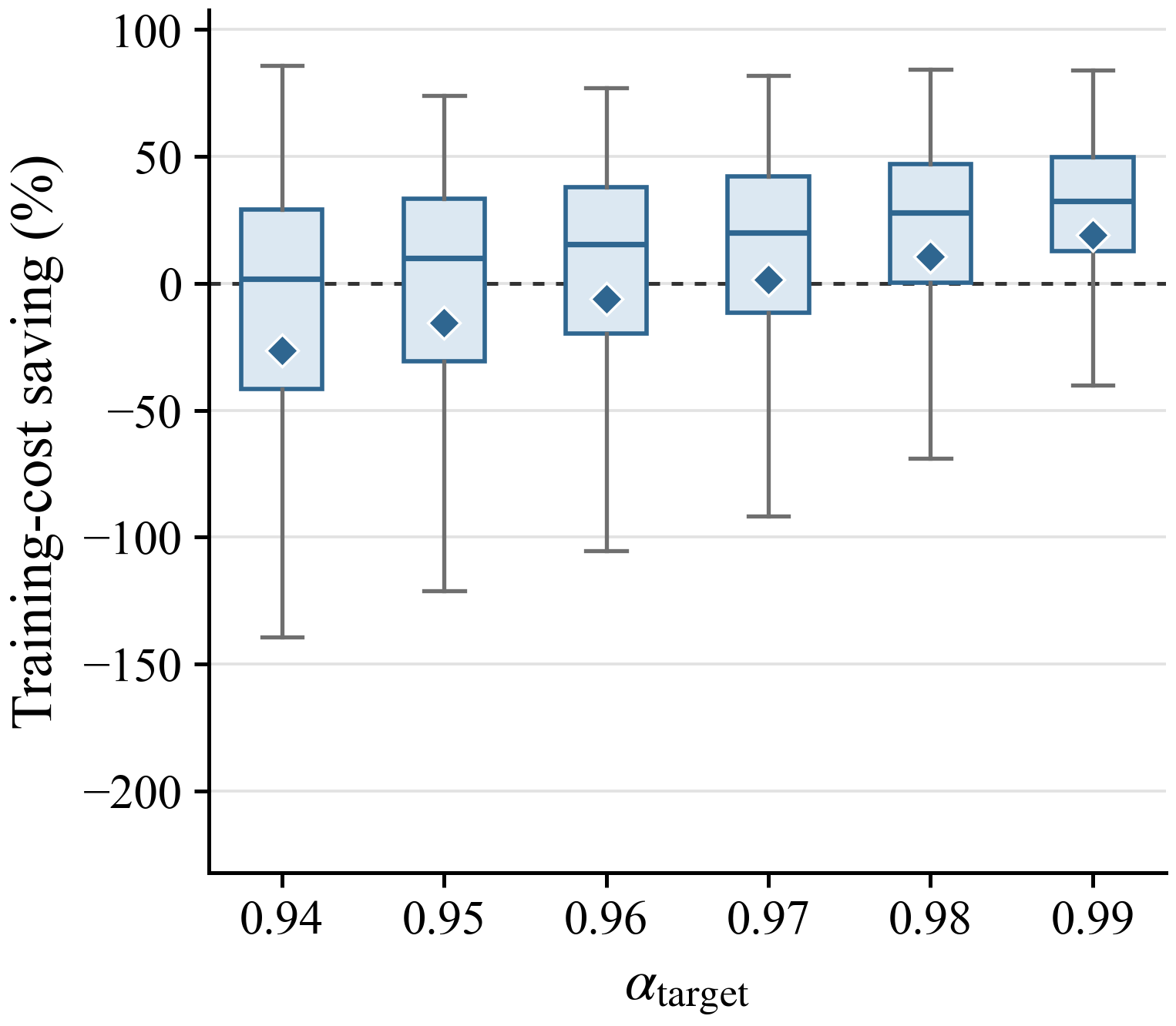}
        \caption{}
        \label{fig:COBYLA_box}
    \end{subfigure}
    \hfill
    \begin{subfigure}[t]{0.554\textwidth}
        \centering
        \includegraphics[width=\linewidth]{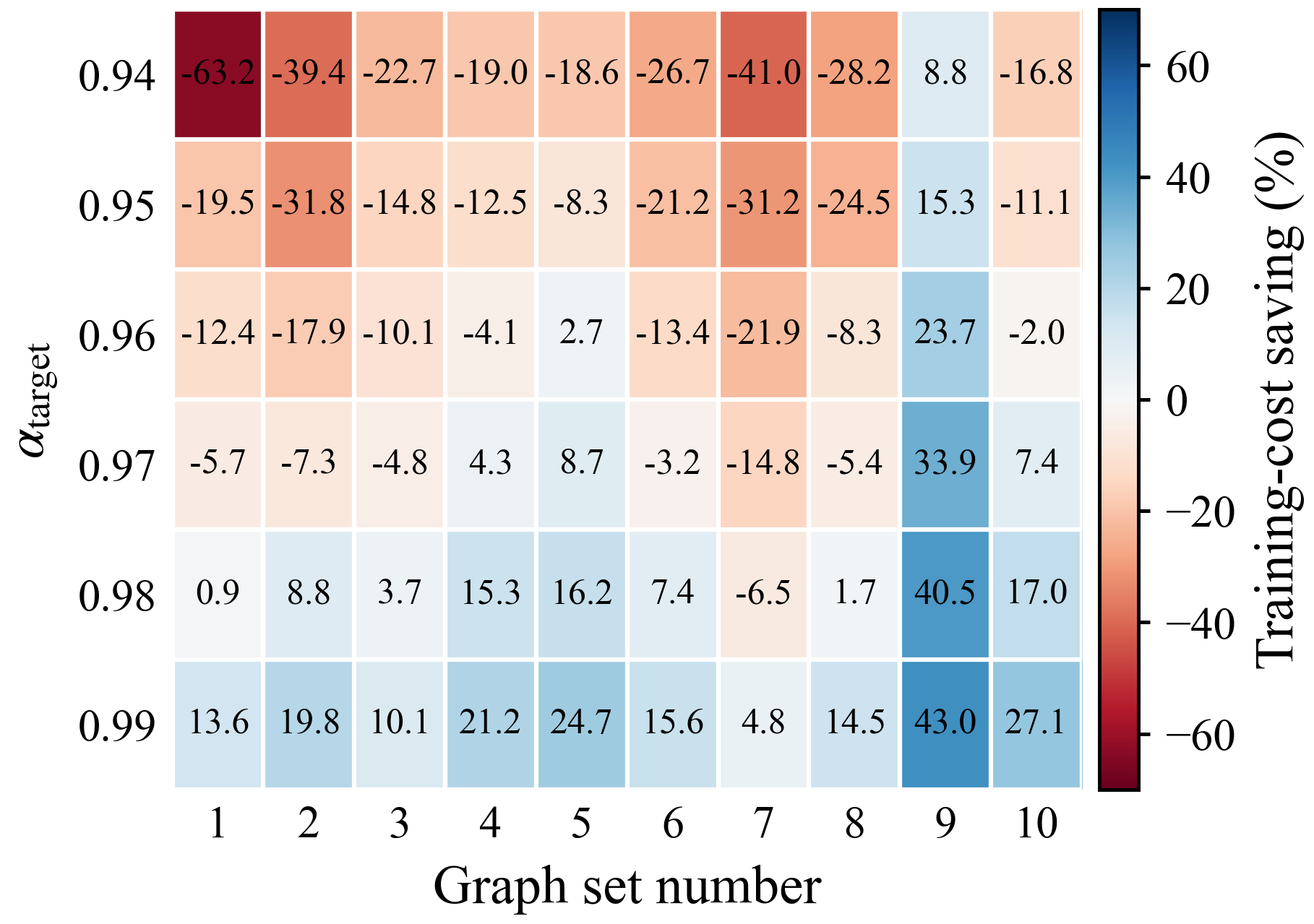}
        \caption{}
        \label{fig:COBYLA_heat}
    \end{subfigure}
    \caption{Percentage savings in training cost achieved by the proposed method compared against COBYLA across varying target approximation ratios $\alpha_{\text{target}}$. The results are presented as (a) box-and-whisker plots detailing the distribution of savings for individual evaluation cases, with diamond markers indicating the mean values, and (b) a heatmap detailing the average savings per graph set.}
    \label{fig:COBYLA}
    \vspace{-3mm}
\end{figure*}

\begin{figure*}[h!]
    \centering
    \begin{subfigure}[t]{0.439\textwidth}
        \centering
        \includegraphics[width=\linewidth]{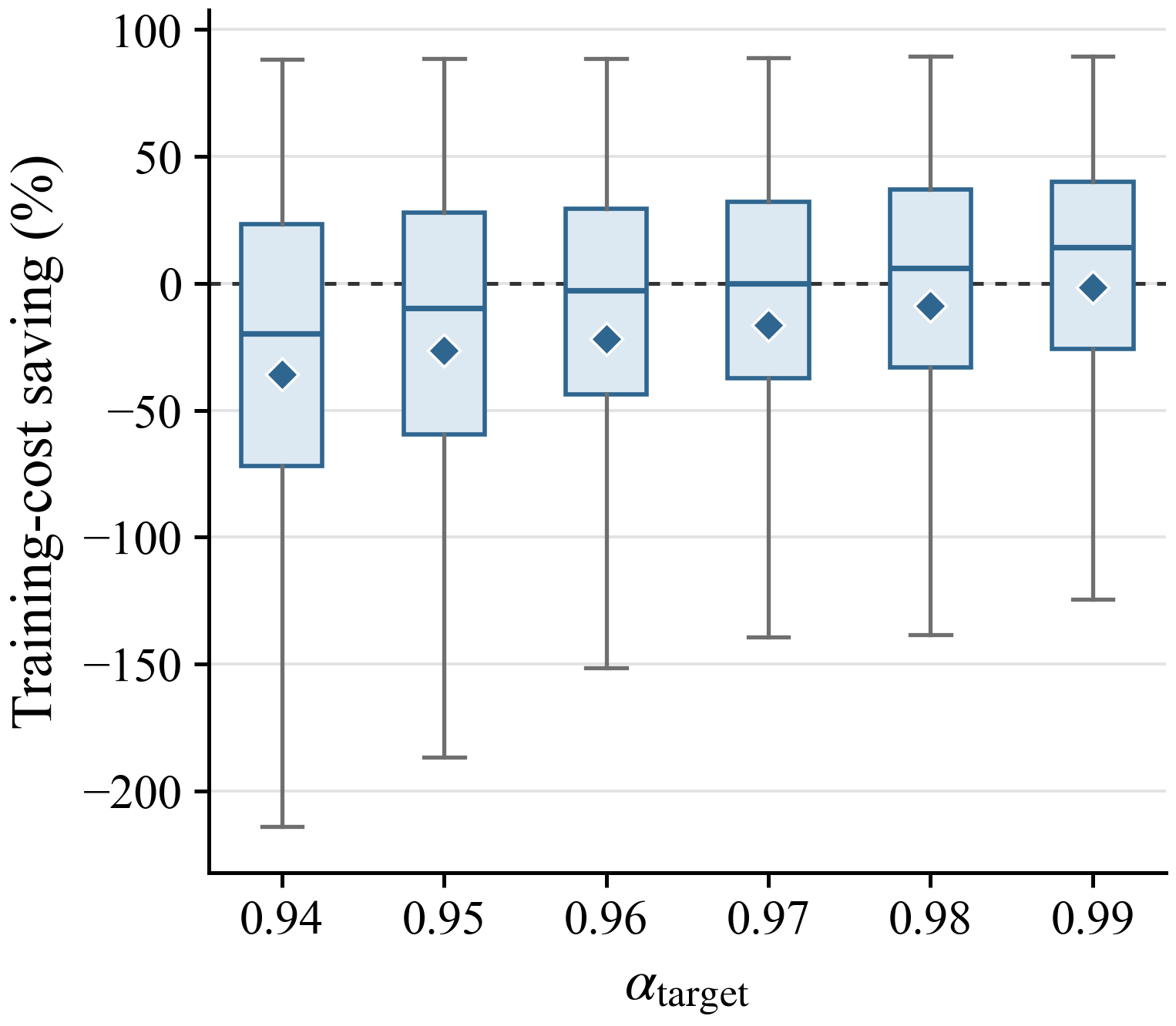}
        \caption{}
        \label{fig:SPSA_box}
    \end{subfigure}
    \hfill
    \begin{subfigure}[t]{0.554\textwidth}
        \centering
        \includegraphics[width=\linewidth]{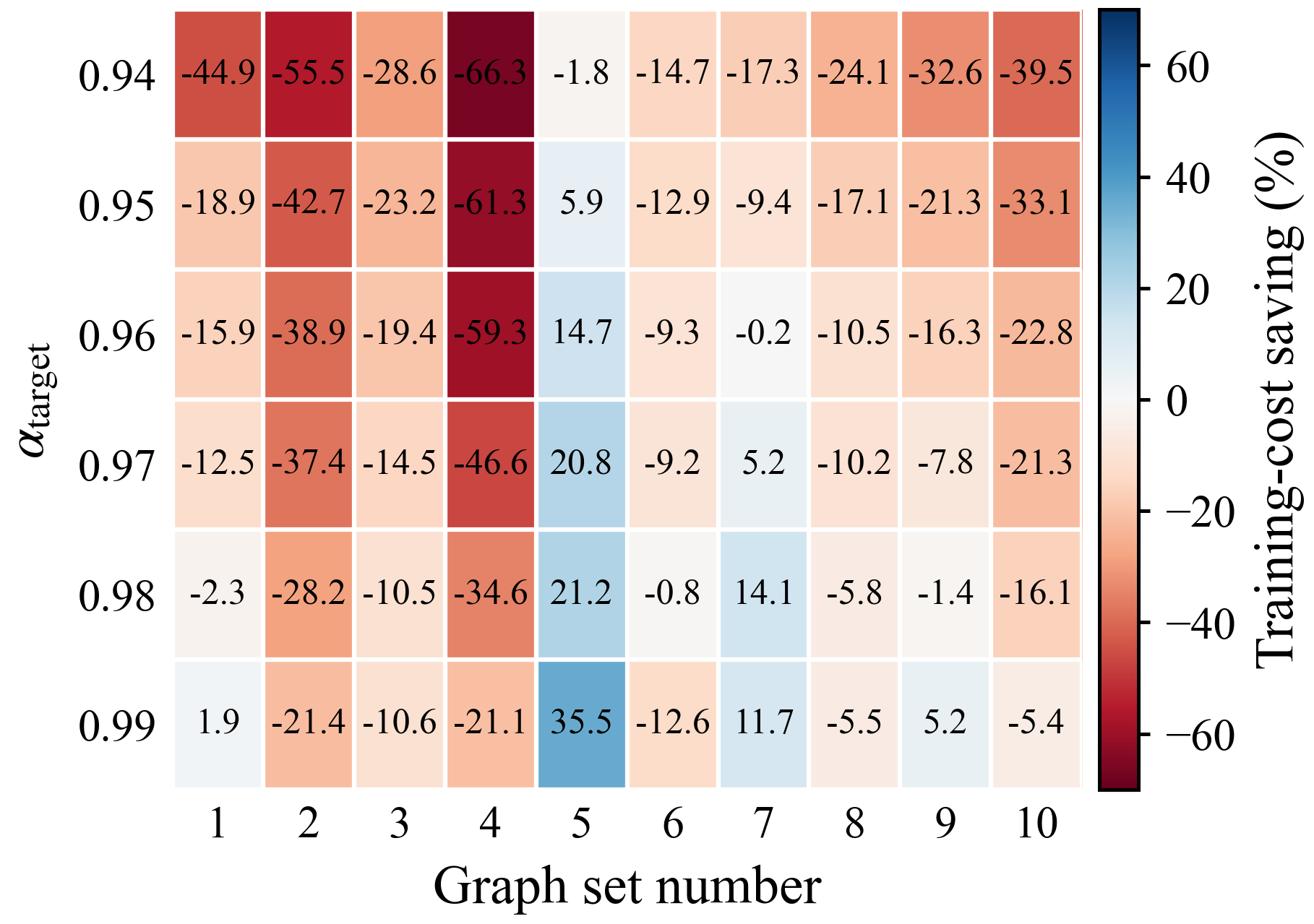}
        \caption{}
        \label{fig:SPSA_heat}
    \end{subfigure}
    \caption{Percentage savings in training cost achieved by the proposed method compared against SPSA across varying target approximation ratios $\alpha_{\text{target}}$. The results are presented as (a) box-and-whisker plots detailing the distribution of savings for individual evaluation cases, with diamond markers indicating the mean values, and (b) a heatmap detailing the average savings per graph set.}
    \label{fig:SPSA}
    \vspace{-3mm}
\end{figure*}

The empirical evaluation demonstrates that the proposed coordinate descent training strategy significantly reduces the computational overhead required to achieve a target approximation ratio. The cost-saving distributions presented in Figs.~\ref{fig:fiveOptimiser}, \ref{fig:COBYLA}, and \ref{fig:SPSA} must be interpreted as strictly conditional comparisons: each plotted value is computed exclusively over instances for which both the proposed method and the corresponding baseline successfully attained the designated target approximation ratio. 

Across the evaluated graph families, the proposed method yields consistently positive and substantial cost reductions when compared against Adam, GD, L-BFGS-B, NM, and Powell at a baseline target approximation ratio of $\alpha_{\text{target}}=0.94$, as visualised in Fig.~\ref{fig:fiveOptimiser}. An analysis of the overall distributions reveals that gradient-based baselines are particularly outperformed, with the proposed method achieving mean savings of 96.2\% against GD and 92.0\% against Adam. While derivative-free methods exhibit a wider variance in their cost-saving distributions, their overall performance still strongly favours the proposed method, demonstrating mean savings of 88.3\% against L-BFGS-B, 80.2\% against NM, and 77.0\% against Powell. This systematic advantage suggests that explicitly exploiting the analytic, coordinate-wise sinusoidal structure of the multi-angle quantum approximate optimisation algorithm (ma-QAOA) objective landscape successfully circumvents the extensive evaluations that generic multidimensional routines expend on estimating descent directions, adapting step sizes, or constructing local surrogate models.

\begin{table}[t!]
\small
\centering
\caption{Graph families used in the numerical experiments.}
\label{tab:graphCases}
\begin{tabular}{c c c | c c c}
\toprule[1pt]
Set & $|V|$ & Edge probability & Set & $|V|$ & Edge probability \\
\midrule
1 & 9  & 0.6 & 6  & 12 & 0.1 \\
2 & 10 & 0.6 & 7  & 12 & 0.1 \\
3 & 11 & 0.6 & 8  & 14 & 0.45 \\
4 & 12 & 0.6 & 9  & 16 & 0.35 \\
5 & 12 & 0.2 & 10 & 18 & 0.30 \\
\bottomrule[1pt]
\end{tabular}
\vspace{-3mm}
\end{table}

\begin{table}[t!]
\small
\centering
\caption{Number of cases where the proposed method and each of baseline optimisation algorithms cross a target approximation ratio of $0.94$, out of 750 total cases.}
\label{tab:rsCrossesOtherCounts}
\begin{tabular}{l | r r r r}
\toprule[1pt]
Baseline & Proposed Only & Baseline Only & Both & Neither \\
\midrule
Adam     & 19  & 17 & 696 & 18 \\
COBYLA   & 11  & 18 & 704 & 17 \\
GD       & 74  & 7  & 641 & 28 \\
L-BFGS-B & 77  & 9  & 638 & 26 \\
NM       & 317 & 5  & 398 & 30 \\
Powell   & 183 & 0  & 532 & 35 \\
SPSA     & 123 & 7  & 592 & 28 \\
\bottomrule[1pt]
\end{tabular}
\vspace{-3mm}
\end{table}

While the conditional cost-saving metrics in Fig.~\ref{fig:fiveOptimiser} illustrate sample efficiency during successful trials, Table~\ref{tab:rsCrossesOtherCounts} provides critical context regarding the convergence reliability and robustness of the evaluated optimisers. Because cost-saving calculations inherently exclude failed runs, they can mask an algorithm's susceptibility to premature convergence. The proposed method demonstrates superior reliability, crossing the $\alpha_{\text{target}}=0.94$ threshold in a significantly higher number of instances where several baselines fail entirely, particularly GD, L-BFGS-B, NM, Powell, and SPSA. 

This discrepancy between conditional efficiency and absolute reliability is most acute for SPSA. Although subsequent conditional analyses in Fig.~\ref{fig:SPSA} suggest SPSA is highly sample-efficient when it converges, Table~\ref{tab:rsCrossesOtherCounts} reveals a substantial robustness deficit. Out of 750 total evaluation cases, there are 123 instances where solely the proposed method crosses the target, compared to a mere 7 instances where only SPSA succeeds. This stark contrast indicates that while SPSA's dimension-independent stochastic gradient approximations may occasionally facilitate rapid initial progress, the method is highly prone to stagnation or convergence to suboptimal local minima. By contrast, the proposed coordinate descent framework offers a substantially more stable and reliable optimisation trajectory.

Because COBYLA and SPSA proved to be the most competitive baselines at the initial threshold, they are analysed across a spectrum of increasingly stringent target ratios in Figs.~\ref{fig:COBYLA} and \ref{fig:SPSA}, motivating a detailed comparison under requirements for higher solution quality. The evaluation against COBYLA reveals a clear dependency on the required approximation ratio. As depicted in Fig.~\ref{fig:COBYLA_box}, the comparative advantage is mixed at lower targets, with mean savings of -26.6\% at $\alpha_{\text{target}}=0.94$, indicating COBYLA often requires fewer evaluations to satisfy relaxed criteria. However, the proposed method demonstrates superior asymptotic scaling as $\alpha_{\text{target}}$ approaches 1, crossing into positive efficacy at $\alpha_{\text{target}}=0.97$ and achieving an overall mean saving of 19.2\% at $\alpha_{\text{target}}=0.99$. The heatmap in Fig.~\ref{fig:COBYLA_heat} corroborates this transition, showing broad positive savings across the majority of graph topologies at the highest fidelity requirements.

Conversely, the detailed breakdown in Fig.~\ref{fig:SPSA} illustrates that SPSA remains a formidable competitor solely in terms of conditional evaluation speed. The cost-saving distributions in Fig.~\ref{fig:SPSA_box} reveal a consistently wide variance, with conditional mean savings gradually shifting from -35.9\% at $\alpha_{\text{target}}=0.94$ to a nearly neutral -1.6\% at $\alpha_{\text{target}}=0.99$. The heatmap Fig.~\ref{fig:SPSA_heat} underscores SPSA's rapid convergence capabilities when successful, showcasing substantial negative savings across multiple configurations. Nevertheless, as established by the crossing counts in Table~\ref{tab:rsCrossesOtherCounts}, this conditional efficiency is achieved at the unacceptable cost of frequent optimisation failures, reaffirming the proposed coordinate descent method as a more robust and scalable solution for high-fidelity ma-QAOA training.

\section{Conclusion}

This paper investigated cyclic coordinate descent for training maQAOA on the maximum independent set problem, a widely studied benchmark in constrained combinatorial optimisation. By exploiting the single frequency dependence of the maQAOA objective on each individual angle, the proposed method replaces generic multidimensional search steps with analytic coordinate updates requiring only a small number of expectation value evaluations. Although we apply this cyclic coordinate descent specifically to maQAOA, the theoretical foundation of the method, namely that the expectation value with respect to a chosen parameter can be analytically expressed as a finite Fourier series, is broadly applicable for optimising parameters across general tasks and problems solved via PQCs. 

Experiments on connected Erdős Rényi graph families demonstrate that this analytic approach yields substantially lower target-reaching training costs and significantly higher convergence reliability than most standard gradient based, stochastic, and derivative free baselines. Specifically, the proposed framework consistently outperformed Adam, GD, LBFGSB, NM, and Powell in both absolute sample efficiency and overall success rate. Furthermore, while algorithms such as COBYLA proved competitive at relaxed approximation thresholds, the coordinate descent method exhibited superior asymptotic scaling, becoming increasingly favourable as the required solution fidelity approached theoretical limits. 

Among the evaluated baselines, SPSA emerged as the strongest comparator in terms of conditional evaluation speed. However, our comprehensive evaluation revealed that the dimension independent stochastic gradient estimates of SPSA come at the severe cost of high optimisation failure rates and a marked susceptibility to premature convergence. In contrast, the proposed coordinate descent method maintained a highly stable and reliable optimisation trajectory across all target ratios. Future work will therefore expand this comparative analysis to include both the baselines evaluated herein and other emerging state of the art methods across larger problem sizes and deeper circuit architectures. Furthermore, extending these evaluations to realistic physical constraints, such as finite shot sampling and hardware noise models, will ascertain whether the pronounced efficiency and reliability trends demonstrated here persist beyond the moderate scale instances considered in this study.


\begingroup
\bibliography{ref}

\bibliographystyle{plainnat}
\endgroup







\end{document}